\RequirePackage{cmap}

\documentclass[conference]{IEEEtran}[2015/08/26]

\usepackage[zerostyle=b,scaled=.75]{newtxtt}
\usepackage{mwe}
\usepackage{authblk}

\usepackage[utf8]{inputenc} 
\usepackage{multicol}

\usepackage{tikz}
\usepackage{pgfplots}
\usepackage[ngerman,main=english]{babel}
\addto\extrasenglish{\languageshorthands{ngerman}\useshorthands{"}}

\usepackage{upquote}

\usepackage{paralist}

\usepackage{csquotes}

\usepackage{url}
\makeatletter
\g@addto@macro{\UrlBreaks}{\UrlOrds}
\makeatother

\usepackage{booktabs}

\usepackage{xcolor}
\usepackage{algcompatible}

\usepackage{pdfcomment}

\ifCLASSOPTIONcompsoc
  \usepackage[%
    square,        
    comma,         
    numbers,       
    sort           
  ]{natbib}
\else
  \usepackage[%
    square,        
    comma,         
    numbers,       
    sort&compress 
  ]{natbib}
\fi

\usepackage{etoolbox}

\usepackage[framemethod=tikz]{mdframed}
\mdfdefinestyle{proofstyle}{%
middlelinewidth=1pt,%
frametitlerule=true,%
frametitlerulewidth=.5pt,
innertopmargin=\topskip,
}

\mdtheorem[style=proofstyle]{proof}{Proof}

\usepackage{arevmath}     
\usepackage[noend]{algpseudocode}
\makeatletter
\patchcmd{\NAT@test}{\else \NAT@nm}{\else \NAT@hyper@{\NAT@nm}}{}{}
\makeatother

\usepackage{xcolor}
\usepackage[linesnumbered,ruled,vlined]{algorithm2e}

\usepackage{hyperref}
\hypersetup{hidelinks,
  colorlinks=true,
  allcolors=black,
  pdfstartview=Fit,
  breaklinks=true}
\usepackage[all]{hypcap}

\usepackage[capitalise,nameinlink]{cleveref}
\crefname{lstlisting}{\lstlistingname}{\lstlistingname}
\Crefname{lstlisting}{Listing}{Listings}

\newenvironment{listing}[1][htbp!]{\begin{figure}[#1]}{\end{figure}}

\usepackage{xspace}

\DeclareFontFamily{U}{MnSymbolC}{}
\DeclareSymbolFont{MnSyC}{U}{MnSymbolC}{m}{n}
\DeclareFontShape{U}{MnSymbolC}{m}{n}{
  <-6>    MnSymbolC5
  <6-7>   MnSymbolC6
  <7-8>   MnSymbolC7
  <8-9>   MnSymbolC8
  <9-10>  MnSymbolC9
  <10-12> MnSymbolC10
  <12->   MnSymbolC12%
}{}
\DeclareMathSymbol{\powerset}{\mathord}{MnSyC}{180}

\usepackage{stfloats}

\SetKwInput{KwInput}{Input}                
\SetKwInput{KwOutput}{Output}              

\begin{document}

\title{Building Graphs at a Large Scale: Union Find Shuffle}
\author{Saigopal Thota}
\author{Mridul Jain}
\author{Nishad Kamat}
\author{Saikiran Malikireddy}
\author{Pruthvi Raj Eranti}
\author{Albin Kuruvilla}
\affil{Walmart Labs \\ \{sthota, mridul.jain, nkamat, saikiranreddy.malikireddy, eranti.raj, albin.kuruvilla\}@walmartlabs.com }

\maketitle

%
%

\begin{abstract}

Large scale graph processing using distributed computing frameworks is becoming pervasive and efficient in the industry. In this work, we present a highly scalable and configurable distributed algorithm for building connected components, called Union Find Shuffle (UFS) with Path Compression. The scale and complexity of the algorithm are a function of the number of partitions into which the data is initially partitioned, and the size of the connected components. We discuss the complexity and the benchmarks compared to similar approaches. We also present current benchmarks of our production system, running on commodity out-of-the-box cloud Hadoop infrastructure, where the algorithm was deployed over a year ago, scaled to around 75 Billion nodes and 60 Billions linkages (and growing). We highlight the key aspects of our algorithm which enable seamless scaling and performance even in the presence of skewed data with large connected components in the size of 10 Billion nodes each. 

\end{abstract}

%
\IEEEpeerreviewmaketitle

\section{Introduction}
\label{sec:intro}
As more data is collected with Internet-based applications, IoT applications, etc., efficient data structures, processing algorithms, and storage paradigms become crucial for managing data. Large-scale graph processing is quite ubiquitous where substantial data can be modeled as a graph eg: knowledge graph, customer identity graph, social graphs, recommendation engines, maps, supply chain etc. 

Depending on the applications, the type of data, and noise in the data, resultant graphs when built can have various characteristics. In our experience, we have come across a mix of different kinds of graphs: (a) Sparse graphs - where each connected component has a small number of nodes and few edges connecting them, (b) Dense graphs - Small graphs connected by a large number of edges among themselves, (c) Graphs forming long chains, and (d) Large connected components (LCC) - where tens of millions of nodes connected together by edges forming one single connected component. It is important to develop algorithms that cater to a variety of graphs, since in many cases, we do not know the distribution of the data and the types of graphs that the data will form.

In this work, we present a highly scalable algorithm that we developed and deployed on commodity Hadoop distribution that works with different kinds of graphs. We currently run production instances of the algorithm to process 75 Billion nodes and 60 Billion edges (and growing) everyday. 

The efficiency of most of the big data processing systems, is largely determined by the implementation of underlying distributed algorithm. Beyond the most obvious strategy of data partitioning, to achieve performance, the most important factors are - shuffle (amount and frequency of data movement), iterations (number of Map-Reduce jobs required), partition size (resources required to store data per partition, and partition process time), and convergence conditions of the overall algorithm. Classic Union Find data structures and algorithms need to account for all these factors in a distributed setting. To that affect, we discuss the performance of our algorithm w.r.t scale, compute cost, and runtime. We discuss how our approach is different from other works in general, and more efficient specifically for dense graphs or even skewed datasets.

In this work, we present the following contributions:
\enumerate
\item{Proposed connected component algorithm - Union Find Shuffle (UFS) with path compression}
\item{Detailed algorithmic bounds and takeaways on how UFS is providing the speed up}
\item{Benchmarks and observations on the performance of UFS against similar algorithms, and different data distributions}
\item{Discussion on related problems and future work}

\section{Related Work}
\label{sec:relatedwork}

When all the edges of a graph fit within a single machine, there are standard methods to generate connected components such as Weighted Union Find~\cite{Union_Find}.  In case of large input data, among distributed approaches to building connected components, there are three popular paradigms in which this problem is approached. They are the parallel RAM Model (PRAM), Bulk synchronous parallel paradigm (BSP) and Map Reduce (MR). 

The PRAM model enables parallel computations using several processors using shared memory as a method of communication and data exchange. There are multiple notable works on building connected components using PRAM with concurrent writes enabled on the common shared memory, presented in~\cite{pram, pram2}. PRAM implementations are more complex to implement and are prone to errors compared to other methods for building connected components using today's distributed computing frameworks.

Bulk synchronous processing approaches, on the other hand, perform a series of parallel generic computation steps called supersteps, followed by synchronization. This step is repeated by passing the output data from the previous step to the next superstep. Some popular graph processing systems such as Giraph and Pregel~\cite{pregel} are based on BSP. We have used GraphX~\cite{GraphX} which is based on Pregel for benchmarking, but in our experience, GraphX does not scale beyond a particular number of edges, and the computation becomes intractable with large connected components. Other works show that BSP has latency constraints compared to Map-Reduce computation~\cite{logrounds}. 

A number of Map-Reduce based algorithms are proposed where iterative map reduce operations result in building connected components. There are methods presented in ~\cite{mr1, mr2}, where the algorithms converge iteratively with iterations in the order of $O(d)$, where d is the diameter of the largest connected component. There are vertex centric approaches that work great with Map Reduce frameworks, where the core computation of the algorithm is decentralized i.e.,  focused around a vertex and its neighbors, followed by subsequent shuffles. This iterative approach leads to a faster termination condition where all nodes are converged to their respective connected components.

There are a few approaches~\cite{vertex_prune1, vertex_prune2} that follow a method called vertex pruning, where some nodes (after reaching the termination condition) are dropped from being carried to further iterations of the algorithm. Vertex pruning helps the algorithm to scale further since the amount of data  shuffled and processed reduces from one iteration to another. Our algorithm, follows an aggressive vertex pruning strategy discussed further. 

There are two contributions that are most relevant to our work. In~\cite{ls-ss}, the authors present two variants: two-phase and alternating algorithms called Large Star and Small Star for generating connected components. In this work, each edge is considered from the point of view of the both the nodes for the shuffle operation, doubling the input data requiring higher distributed memory. The approach is not scalable on noisy graphs with skewed distribution of the degree of nodes. The scalability and performance of the algorithm is compared in section~\ref{benchmarks} .

Cracker is another algorithm proposed in~\cite{cracker}, where the authors present a vertex-centric approach similar to our work. In this algorithm, the method is to construct a tree iteratively, given the edges of the graph. This work also discusses vertex pruning. The first few iterations will be heavy to process since all input edges are considered from the perspective of both the nodes in the edge to find the minimum neighbor.  The vertex pruning strategy presented in the paper is iterative in a Breadth First manner.

Our algorithm differs from existing works in three core aspects: 
\begin{enumerate}
\item
Configurability in the first phase of the algorithm, where the input data is divided into partitions (\emph{k}) based on parallel compute bandwidth available. This determines the trade-off of cost vs. parallelism and speedup. The number of partitions also help in providing the scale. On each partition, a local Union Find algorithm with path compression is run to create connected components.
\item
In the second phase, the algorithm creates a minimum spanning tree of the connected components, and provides the flexibility of running an optional path compression step (phase three) to create a star-graph. This flexibility enables users to leverage lazy path compression techniques, where the final parent node is discovered in an amortized fashion.
\item
Vertex pruning strategy is aggressive in our algorithm, since our target is to create a minimum-spanning tree in the second phase as opposed to a star-graph. This results in a steep decrease in the data to be processed with iterations in the shuffle steps. 
\end{enumerate}
With these strategies, our algorithm scales gracefully even with the presence of noisy and skewed data such as one connected component consisting of 10 Billion nodes. This is discussed further in the benchmarks section.

\section{Union Find and Disjoint Sets}
\label{sec:uf}
\begin{figure*}[htb!]
  \centering
  \begin{minipage}[b]{0.33\textwidth}
    \fbox{\includegraphics[scale=0.4]{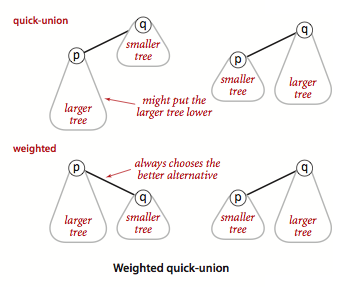}}
    \caption{Weighted Union~\cite{Union_Find}}
    \label{fig:fig1}
  \end{minipage}
  \hfill
  \begin{minipage}[b]{0.33\textwidth}
    \fbox{\includegraphics[scale=0.1]{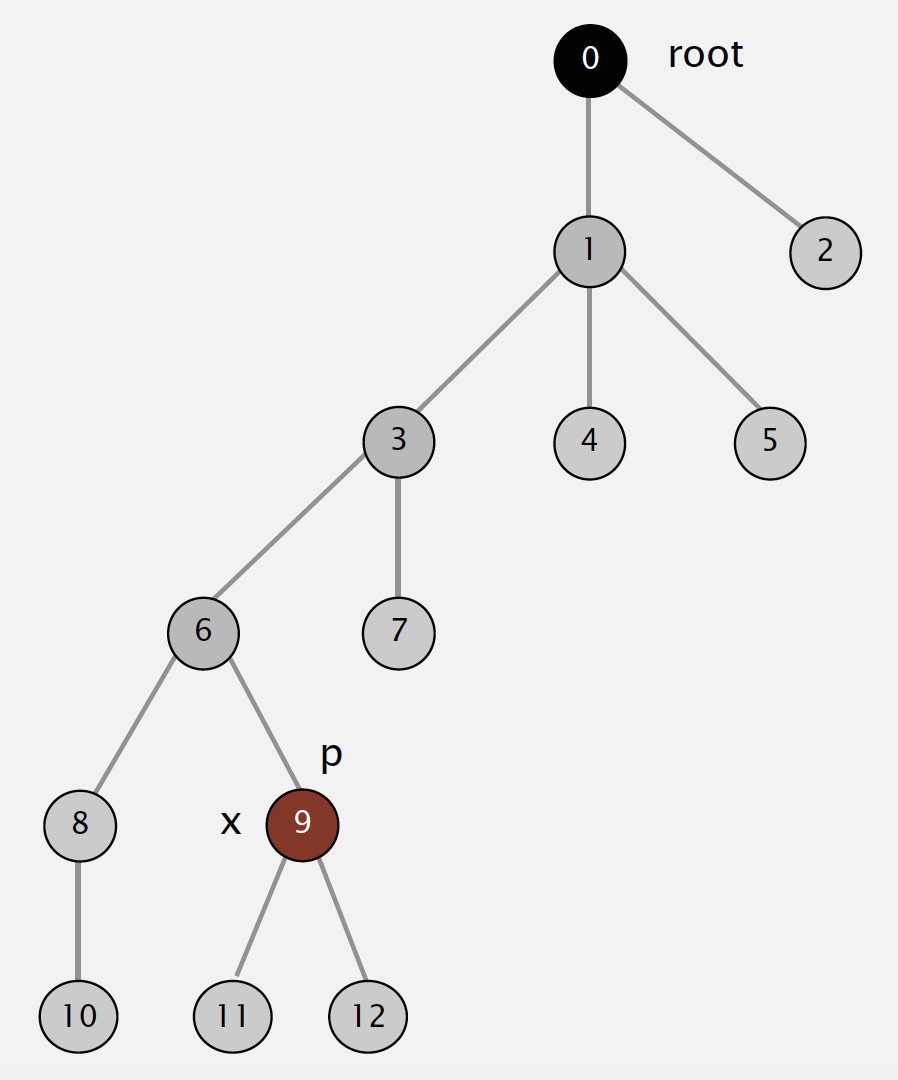}}
    \caption{Before Path Compression~\cite{Union_Find}}
    \label{fig:fig2}
  \end{minipage}
  \hfill
  \begin{minipage}[b]{0.323\textwidth}
    \fbox{\includegraphics[scale=0.1]{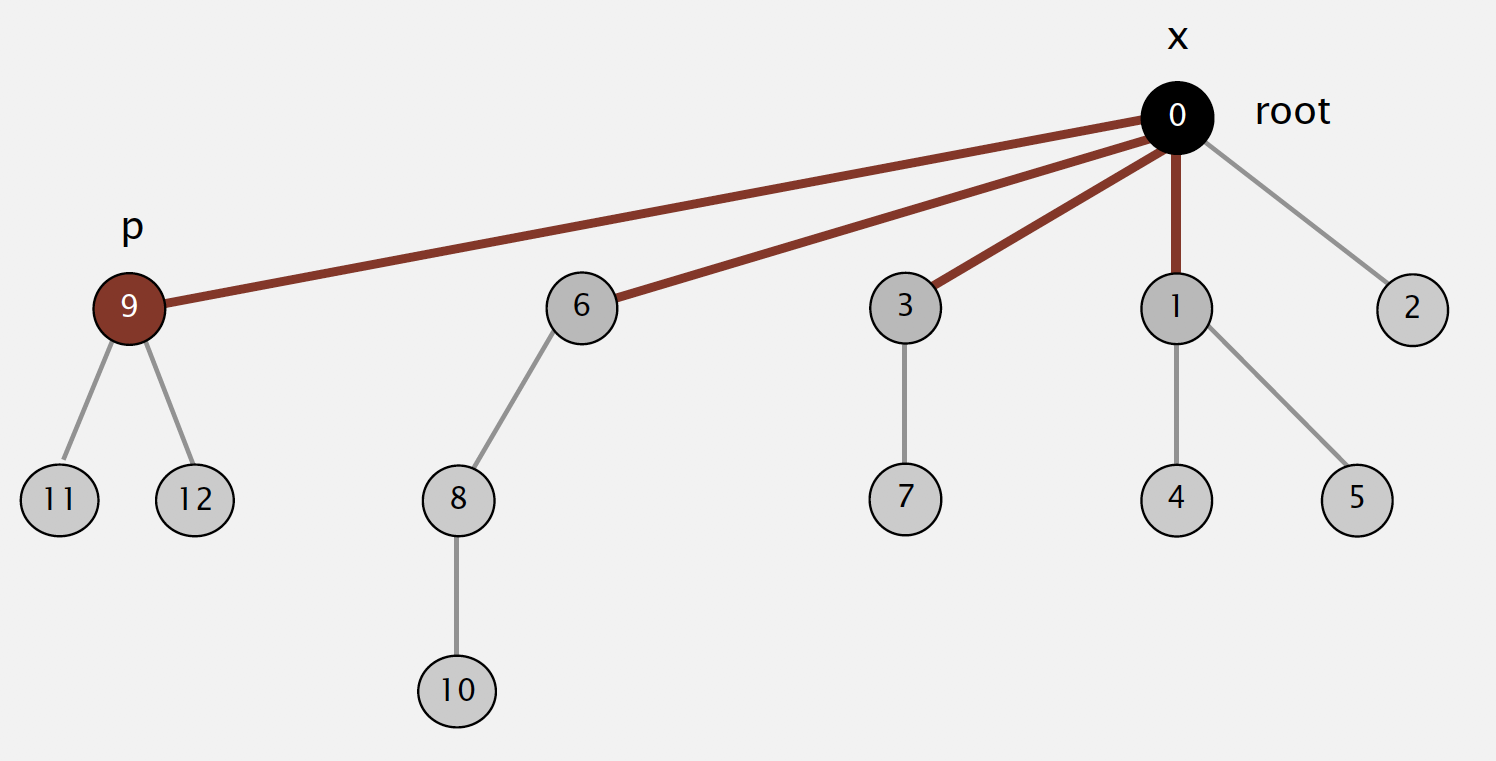}}
    \caption{After Path Compression~\cite{Union_Find}}
    \label{fig:fig3}
  \end{minipage}
\end{figure*}

A connected component is a subgraph in which any two vertices are connected to each other by paths~\cite{cc_def}. Union Find algorithms used on disjoint sets is common way to generate connected components.  The input to any such algorithm is a set of edges, and the goal is to identify all subgraphs formed. The output of the algorithm is in the form of a child-parent relation where the parent node ID represents the subgraph ID, and all nodes that belong together in a subgraph get the same parent node.

Connected component algorithm Weighted Union Find consists of two major steps:
\begin{itemize}
\item{Weighted Quick Union: When a new edge is seen that results in two connected components merging together, this step ensures that the depth of any node, x is at most log N, where N is the number of nodes in the tree. It achieves this by keeping track of size of each tree and maintaining a balance by linking root of smaller tree to root of the larger tree, whenever a union is performed between the given nodes of relevant trees (Fig.~\ref{fig:fig1})}.
\item{Path Compression: Path compression helps to further flatten the tree, by making the children point to the root node directly, at amortized cost, when finding the root for a given child. Hopcroft-Ulman, Tarjan analyzed Weighted Quick Union with Path Compression and found that any sequence of M union-find operations on N objects makes <= $O(N+M*log(N))$ array accesses. In practice this turns out to be almost linear in time proportional to N (Figs~\ref{fig:fig2} and \ref{fig:fig3})}.
\end{itemize}

\section{Union Find Shuffle with Path Compression}
Union Find Shuffle (UFS) with Path Compression algorithm is a distributed algorithm that generates connected components in three phases.

\begin{figure*}[htb!]
\centering
  \fbox{\includegraphics[scale=0.6]{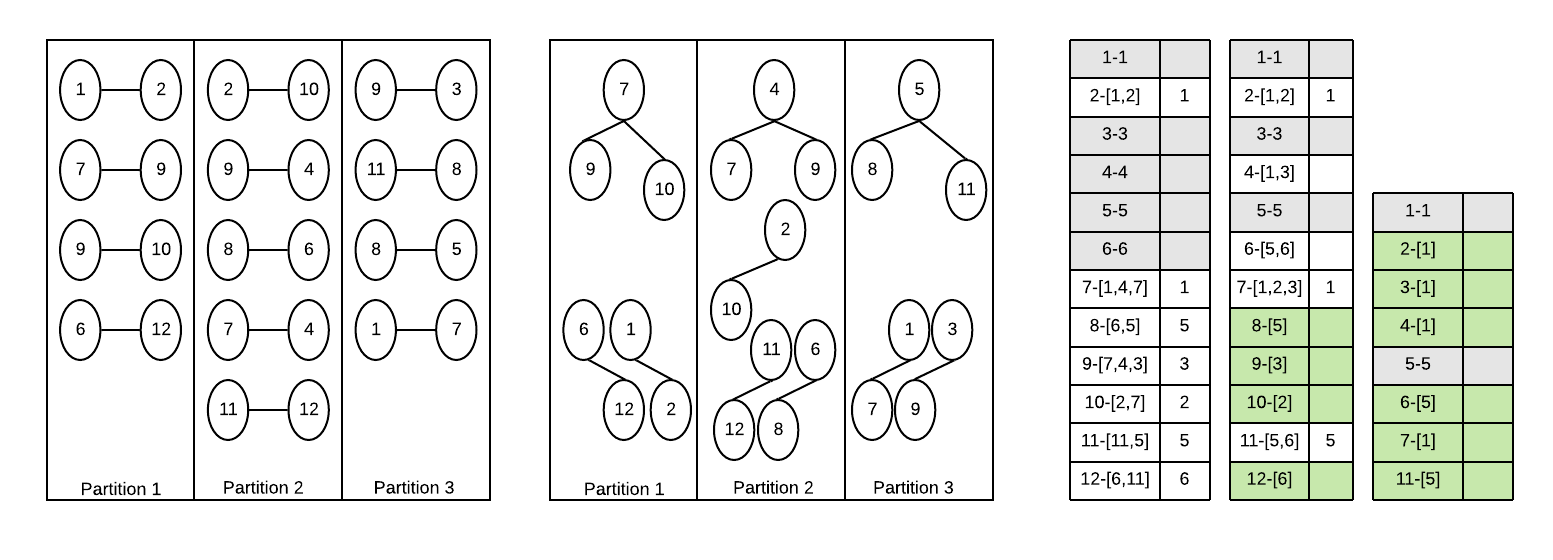}}
  \caption{LoadPartitions->WeightedUnion->ProcessPartition (iterate till convergence)}
  \label{fig:fig4}
\end{figure*}


\begin{algorithm*}
\caption{Union Find Shuffle} 
\scriptsize
\SetKwInput{KwInput}{Input}                
\SetKwInput{KwOutput}{Output}              
\DontPrintSemicolon
  
  \KwInput{input - part files}
  \SetKwFunction{FMain}{Main}
  \SetKwFunction{FWeightedUnion}{WeightedUnion}
  \SetKwFunction{FProcessPartition}{ProcessPartition}
  \SetKwFunction{FPathCompression}{PathCompression}
    \SetKwFunction{FSelfJoin}{SelfJoin}
 
  \SetKwProg{Fn}{}{:}{}
  \Fn{\FWeightedUnion{$partition$}}{
    Let np = $\{\}$ \tcp*{Initialize sets i.e local partition state}
    \For{l in partition} {  
      u, v = l[0], l[1]\;
      p(u) = FindWithPathCompression(u) \tcp*{Classic Union Find with Path Compression}
      p(v) = FindWithPathCompression(v) \tcp*{p(u),p(v) - parents of u, v resp}
      \If {p(u) == p(v)} {
        continue\;
      }
                     
      \If {p(u).treesize >= p(v).treesize} {    \tcp*{Apply weighted union} 
        p(u).treesize+=p(v).treesize            \tcp*{i.e attach small tree to large tree} 
        np+=p(u)\;
        ShuffleEmit(v,p(u)) \tcp*{Shuffle by left key i.e group/partition by child} \label{WU:se1}
      }
      \Else {
        p(v).treesize+=p(u).treesize\;
        np+=p(v)\;
        ShuffleEmit(u,p(v)) \tcp*{Shuffle by left key i.e group/partition by child} \label{WU:se2}
      }
    }
    \For{parent in np} {
      ShuffleEmit(parent,parent) \tcp*{Emit if you are a NewParent} \label{WU:se3}
     }
  }
  \SetKwProg{Fn}{}{:}{}
  \Fn{\FProcessPartition{$partition$}}{
    Let cp=$\{\}$\;
    \For{l in partition}{
      cp+=l[1]\;  
    }
    \If {cp.size == 1} {
      \If {l[0]==l[1]}{ 
        \KwRet \tcp*{Ignore all parent-parent linkages} \label{PP:ppl}
      }
      \tcp{Terminal condition reached: found a unique parent, so Shuffle by left key i.e partition by parent and forward to checkpoint}
      \KwRet Checkpoint(cp, l[0]) \label{PP:se2d} 
    }
    np=min(cp) \tcp*{Choose the smallest node as the parent}
    \For{n in cp}{
      ShuffleEmit(n, np) \tcp*{Shuffle by left key i.e group/partition by child} \label{PP:se}
    }
    ShuffleEmit(l[0],np)
  }
  \;
  
  \SetKwProg{Fn}{}{:}{}
  \Fn{\FSelfJoin($partition$)}{
    \For{l in partition}{
      ShuffleEmit(l[0],l[1]); ShuffleEmit(l[1],l[0])\;
    }
  }

  \SetKwProg{Fn}{}{:}{}
  \Fn{\FPathCompression{$partition$}}{
    Let cc=$\{\}$\;
    Let minnode = min(partition) \tcp*{Find the min node in the partition}
    Let pruned = True\;
    \For{l in partition}{
      \If {l[0] > minnode \&\& l[1] > minnode} {
        pruned = False\;
      }
      cc+=l[0];  cc+=l[1]\;
    }
    
    \If {pruned} {
      \KwRet CheckPoint(partition)\;       \tcp*{Terminal condition reached: minnode is parent, so partition can be pruned}
    }
    \For{n in cc}{
      \tcp*{Group by both nodes}
      ShuffleEmit(minnode, n);     ShuffleEmit(n, minnode)
  }}

  \SetKwProg{Fn}{}{:}{\KwRet}
  \Fn{\FMain}{
    partitions=LoadPartitions(input)
    \tcp*{WeightedUnion fn. runs in parallel on all partitions}
    \tcp*{WeightedUnion function doesn't return a partition. ufpartitions are generated due to Shuffle on WeightedUnion output}
    ufpartitions=WeightedUnion(partitions)\; 
    
    \While{Size(ufpartitions)!=0}
          {
            \tcp*{ufpartitions and output partitions generated due to Shuffle inside ProcessPartition}
            output, ufpartitions = ProcessPartition(ufpartitions)\; 
            WriteToHDFS(output) \tcp*{Write all output partitions to disk intermittently}
          }
    output = SelfJoin(output)\;  \tcp*{Load data from HDFS into partitions and emit an outer self-join}
    \While{Size(output) != 0}
     {
      compressed, output = PathCompression(output)\;
      WriteToHDFS(compressed)\;      
    }    
    \KwRet 0\;
  }
\end{algorithm*}

The input to the algorithm is a set of undirected edges, divided into a set of partitions (depending on the parallel resources available), with roughly equal number of edges. The algorithm (shown below) starts at the Function Main, which initiates the steps of the distributed algorithm. 

\subsection{Local Weighted Union with Path Compression}
The first phase of UFS is a weighted union with path compression step, run on each partition of input data in parallel. This is the key to horizontally scaling the algorithm. Logically, a partition has 1:1 mapping with the VM process, memory boundary; though the underlying framework (map-reduce, Spark), could map a number of partitions to the same executor/process. 

Weighted union find with path compression (as described in section~\ref{sec:uf}) helps in creating connected components, based on the knowledge of linkages within the partition. This is a critical first step which differs from the Large-star, Small-star algorithm in~\cite{ls-ss}.  Running a local Union Find on each partition reduces the volume of data to be shuffled significantly, and helps in scaling the algorithm to billions of linkages. The number of partitions into which the initial data is divided is governed by the amount of parallel processing available to compute local Union Find. The output of this step is in the form of a `child node-parent node' relationship. 

\subsection{Shuffle}
In the second phase, we run a series of shuffle steps iteratively until every node in the subgraph unambiguously identifies a single node as its parent. With this step we make sure that there is no fragmentation among linkages belonging to the same connected component. By the end of stage 2, the graphs formed are minimum spanning trees, where each node has a path to every node in the subgraph. The two key steps of shuffle iterations are presented here.

\subsubsection{ShuffleEmit}

ShuffleEmit is a reduction operation where the output of the previous step is grouped by the child node. In line ~\ref{WU:se1}, ~\ref{WU:se2},~\ref{PP:se} the grouping keys are all the child nodes. If the child was part of multiple partitions in the previous step due to random partition splits, they can be localized into a single new partition to help elect a single parent in the current iteration. Line ~\ref{WU:se3} ensures that any node which became a new parent anywhere across the system, also gets a chance to stand in the election of the parent, in the corresponding partition. ShuffleEmit is called first at the end of the local weighted union with path compression step, and called iteratively with each iteration of the Union Find Shuffle algorithm. 

\subsubsection{ProcessPartition}
Once the new partitions are formed, three types of records show up. (a) If a child node has itself as its parent (called self-linkage), ignore them. These are the grey nodes in figure ~\ref{fig:fig4} (b) If a node has only one parent node after the shuffleEmit step, save it for checkpointing to the memory i.e line~\ref{PP:se2d}. This is known as a termination condition for a node as it found an unambiguous parent (before path compression). This process is also known as vertex pruning. These are the green nodes in figure ~\ref{fig:fig4}.
On the other hand if a child has more than one parent, then choose the node with the minimum or maximum (or some fixed criteria for parent election) number as the parent and make every node in the record a child of this elected parent, and call ShuffleEmit on the data.

\subsubsection{PathCompression}
 Finally, in the third phase, once the shuffle stage of UFS is completed, an optional path compression can be computed on the output of UFS to generate connected components in the form of star-graphs. In practice, it is as simple as a Hive outer self join between the output produced. Since this process can be distributed across multiple machines, this algorithm is horizontally scalable and is highly suitable for distributed computing environments. 

 \subsection {Key Takeaways}

The key takeaways from the algorithm and how different steps help in better scaling, convergence, and performance while building connected components are presented here.
\begin{enumerate}
\item
The number of initial partitions ($k$) for Local Union Find  with path compression provides the power of horizontal scaling where depending on compute available, we can process more data in parallel using more computes.

\item
The shuffle part of the algorithm converges at a complexity rate proportional to the size (number of nodes) in the largest connected component ($S$) in the worst case. It converges in the order of log($S$).
 
 \item
 Aggressive vertex pruning based on our termination condition helps scale the algorithm since less overhead or data is carried forward in further iterations. 
 
 \item
 If a node (\emph{n}) becomes a parent node in some partition, self-loop linkage (\emph{n-n}) is emitted so that in the next iteration \emph{n} knows that it has become a parent for some other node. If n has not become a parent node for other node, it will reach its termination condition, either in this iteration or the next one. Therefore the self-loop linkage facilitates asynchronous decision making that makes the algorithm convergence despite of being distributed completely (meaning no central coordination is required, which helps with true horizontal scalability).

\end{enumerate}

\subsubsection{Benefits of doing Local Union Find with Path Compression}
\begin{enumerate}
\item
The parent child relationship output from Local Union Find eliminates the need to send the data from the perspective of both the nodes of an edge as part of the initial shuffle, which is typically done in most other algorithms. This reduces the shuffle data volume by at least 50\%. 
\item
Local Union Find followed by path compression creates connected components locally which resolves densely connected edges, long chains into a star graph. This reduces shuffle data further. To put this intuitively, Local Union Find with Path compression acts like a combiner in Map Reduce paradigm, creating an aggregate intermediate state which reduces data shuffle.

\item
Local path compression reduces a lot of path compression steps during the final stage (explicit global path compression). High density graphs become flat.

\item
Due to the steps above the cost and time to do first iteration of shuffle reduces significantly. First iteration as we know is the long pole here (due to huge volume of edges) and any efficiency gain in the amount of data shuffled in the first iteration is a big gain.                                                                                  
\end{enumerate}
Hence, leveraging all the key configurations and optimizations, our Union Find Shuffle algorithm scales very well for all kinds of graph data. 
\section{Complexity Analysis}

\begin{table}[tbh!]

\begin{proof*}[Complexity Analysis] \label{ufsproof}

Number of Edges = $N$ \\
    Number of Initial Partitions  = $K$ \\
    Maximum degree of a node = $m (<<N)$ \\
    Maximum size of a connected component = $S$ \\
    From a node's point of view, after the local Union Find: 
    \begin{enumerate}
    \item
Any node $n$ can have a maximum of $k$ parents. \\
    \item
In one extreme case, if $n$ became the parent in all partitions, then it does not have any other parent node and reached the termination condition.\\
    \item 
In another extreme case, if a node's degree is 1, and it is a child of another node, \emph{n'}, it has only one parent node \emph{n'}, and it reached the termination condition.\\
    \item
If a third extreme case, $n$ did not become a parent node in any partition, then it will have a maximum of $p$ parents where ($p<=k$), and n will pick a parent in the current iteration.\\
    \item
 In case, $n$ becomes a parent node in $p$ partitions, and a child node in up to $k-p$ partitions, in the next iteration, node $n$ will have up to a maximum of $S$ child nodes (because of the size of CC) \\
    \item
    When $n$ has $S$ children, $n$ be competing against other parent nodes of the $S$ children, to see if n is chosen as parent or n will find a new parent. \\
    \item
 But, on an average, if we assume that in $50\%$ of cases, $n$ becomes a child node in the current iteration, the number of parent nodes to resolve in next iteration will be reduced to $(S)/2$, eventually reaching 1 parent \\
    \item
 Thus UFS converges in the order of $log(S)$, the size of the largest connected component in the worst case.  \\

    \end{enumerate}

\end{proof*}

\label{table: proof}
\caption{}
\end{table}

The complexity analysis of the UFS algorithm is described here. If the number of input edges are \emph{N}, and if we choose a setup with k partitions to run local union find in parallel, the local union find takes a logarithmic bound in the order of $log(n/k)$, since $n/k$ is the average number of edges per partition. 
After local union find, the shuffle iterations of UFS begin. At the end of the local union find, three extreme scenarios can happen where a given node reached its termination condition and is subject to vertex pruning, and not considered further, it is explained in the proof presented in the table above. In case a given node becomes parent in few partitions, and a child in few partitions, there is a chance to continue to further iterations to figure out what is the best parent node. Therefore, if a node \emph{n} has become parent in some partitions, and child in some partitions, \emph{n}, in the worst case,  will have \emph{S} children, where S is the size of the connected component to which \emph{n} belongs. While participating in the parent selection of each of these \emph{S} children, \emph{n} will have a potential of upto \emph{S+1} parents to choose from, in the next iteration. The worst case scenario for the algorithm to converge slower is when in 50\% of the cases \emph{n} becomes parent and child in the other 50\%. In that case, \emph{n} will have \emph{(S+1)/2} parents to choose from in the next iteration. Thus our algorithm converges with an upper bound on the log(S).  

\section{Benchmarks\label{benchmarks}}
\subsection{Experiment Setup}
To demonstrate the scalability and the run time of the UFS algorithm, we developed experiments to benchmark UFS along with Large star-Small star algorithm and GraphX. The nuances of Large star-Small star are described in Section \ref{sec:relatedwork}. GraphX is the out-of-the-box implementation in Apache Spark for building graph connected components. For benchmarking the performance and scalability of these algorithms, we performed the experiments on a distributed Hadoop cluster in Google Cloud with fixed memory and CPU, while varying the number of input linkages. The compute resources utilized to perform the benchmark experiments with different algorithms across data sizes are presented in Table \ref{tab:resources}.

The goal of the experiment is to measure the feasibility and the duration for the graph to converge among the three algorithms, given the same resources. The data used for the experiment is real data from retail domain with built-in noisy linkages such as high cardinality nodes and edges that create long chain paths, leading to LCCs. We define an LCC as a connected component with more than 5000 nodes for this experiment. The presence of an LCC will make a difference in the performance of the models, since a LCC leads to data skewness if all nodes/edges belonging to a connected component are shuffled to the same node for a reduce operation. It might lead to bottlenecks and in some cases infeasible to converge. To show the importance of Local Union Find, we conducted two experiments, one with and one without doing the local union find. The shuffle operations of the algorithm follow the same steps in both approaches. They are presented as UFS and UFS w/o Local UF respectively. 
\begin{table}[htb!]
  \centering
    \begin{tabular}{|l|r|r|r|}
    \hline
      \textbf{Num of Edges} &  \textbf{Executors} &  \textbf{Executor Memory} & \textbf{Driver Memory}\\ 
      \hline
      1 M & 100 & 4G & 4G\\ 
      10 M & 100 & 4G & 4G\\ 
      100 M & 100 & 8G & 4G\\ 
      1 B & 200 & 16G & 8G\\ 
      12 B & 200 & 32G & 8G\\ 
      43 B & 400 & 32G & 16G\\
      \hline

    \end{tabular}
    \caption {Experiment environment details}
          \label{tab:resources}
\end{table}

\begin{table*}[htb!]
  \centering
    \caption{Benchmarks evaluated across multiple Connected Component Algorithm}
    \label{tab:benchmarks}

    \begin{tabular}{|l|r|r|r|r|r|r|r|}
        \hline
      \textbf{Algo} & \textbf{1M} & \textbf{10M} & \textbf{100M} & \textbf{1B} & \textbf{12B} & \textbf{43B}& \textbf{100B}\\ 
      \hline
      Large star-Small star & 5.2 mins & 7.45 mins & 12 mins & 63.8 mins & 4.6 hrs & N/A& N/A\\ 
      UFS w/o Local UF & 3.1 mins & 6.41 mins & 10.41 mins & 20.35 mins & 2 hrs & 3.65 hrs & 5.9 hrs\\ 
      UFS & 2.4 mins & 4.93 mins & 8 mins & 15.6 mins & 1.54 hrs & 2.8 hrs & 4.2 hrs\\ 
      GraphX & 1.2 mins & 1.4 mins & 7.8 mins & 59 mins & N/A & N/A & N/A\\ 
      \hline
    \end{tabular}
   
\end{table*}

\subsection{Results and Observations}
\begin{figure}[htb!]
  \begin{center}
    \begin{tikzpicture}
      \begin{axis}[
          xmode=log,
          ymode=log,
          width=\linewidth, 
          grid=major, 
          grid style={dashed,gray!30}, 
          xlabel=Number of edges (in Millions), 
          ylabel=Duration (in minutes),
          legend style={at={(0.5,-0.2)},anchor=north}, 
          x tick label style={rotate=45,anchor=east} 
        ]
        \addplot   
        table[x=column 1,y=column 2,col sep=comma] {table.csv}; 
          \addplot 
        table[x=column 1,y=column 2,col sep=comma] {table1.csv}; 
         \addplot 
        table[x=column 1,y=column 2,col sep=comma] {table3.csv}; 
        \addplot 
        table[x=column 1,y=column 2,col sep=comma] {table2.csv};      
        \legend{Large star-Small star, UFS w/o Local UF, UFS, GraphX}
      \end{axis}
    \end{tikzpicture}
    \caption{Benchmark results comparing various algorithms}
     \label{fig:graph_plot}
  \end{center}
 
\end{figure}
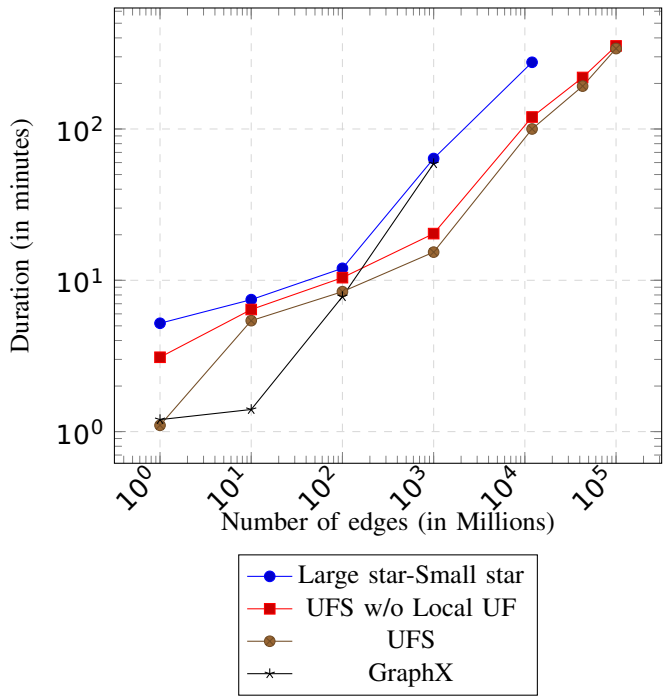

The results of the benchmark experimented are presented in Figure~\ref{fig:graph_plot}. The figure shows the benchmark results, with respect to the duration (in minutes) each of the four algorithms, Large star Small star, UFS w/o Local UF, UFS, and GraphX. The results show the performance on different sized data sets. The data size (in millions) is presented on X-axis and duration (in minutes) is captured on Y-axis. 

The first observation is, for smaller sized input edges, GraphX performs better than UFS and Large Star, but as the data size grows, the duration increases rather sharply before giving up just above 1 billion edges. GraphX has a scaling problem. We show the run time in this case as N/A. 
The second observation is that, Large star-Small star algorithm starts with a duration higher than UFS and GraphX, but after an inflection point around 10 millions the duration shoots up steeply, but performs better than GraphX in scale to around 10 Billion edges. But after 10 Billion, we see that the algorithm does not converge and it fails (shown as N/A). We  In case of the UFS algorithm, the run-time is lower than Large star-Small star for all data sizes, but UFS algorithm seamlessly scales to 100B edges. The time taken for local Union Find is benchmarked earlier at 2.5 minutes to process 40M linkages. But for the overall data of 43 Billion, using multiple executors to run parallel instances of Local Union Find, the processing time is around 8 minutes. This additional time for Local Union Find is giving a significant boost in the runtime performance of overall UFS algorithm and is helping in scaling the algorithm to 100 Billion seamlessly. In all cases, the performance of UFS is better with Local UF than in the case without Local UF. Accurate values of the input number of edges, and the corresponding runtimes for all the algorithm are presented in  Table \ref{tab:benchmarks}.

\section{Conclusion}

In conclusion, the Union Find Shuffle algorithm presented in this work provides a scalable method to build connected components on graphs with different characteristics in terms of size, density, skewness, etc.  The flexibility in the choice (a) between compute cost and the time to build connected components, and (b) build the minimum spanning tree vs. star graph are some novel features of the algorithm. Benchmarks presented in the paper conclude that the algorithm can be readily implemented for building graph connected components using commodity hardware with Hadoop infrastructure. The comparison of our algorithm with other similar methods show that our algorithm has a competitive runtime as well as scale. 

Distributed Computation of graph traversals at scale and  traversing large connected components in the order of 10 billion nodes, etc., are some of the challenging problems that are extensions to the current work. In addition to running graph algorithms in a distributed processing frameworks such as Spark, we are currently working on building connected components in real-time on streaming data sets. For these algorithms, there are challenges in terms of scale, concurrency, communication (synchronous vs. asynchronous),  that we are working on currently.

\bibliographystyle{IEEEtranN} 

\bibliography{paper}

\end{document}